\documentclass[
showpacs
,preprint%
,prl,aps]{revtex4}
\usepackage{graphicx}
\usepackage{psfrag}
\usepackage{epic,epsfig}
\usepackage[dvips]{color}
\begin{document}
\title{The Stefan-Boltzmann law in a small box\\ and the pressure deficit
in hot SU(N) lattice gauge theory}
\author{Ferdinando Gliozzi }
\affiliation{
Dipartimento di Fisica Teorica, Universit\`a di Torino and\\ INFN,
Sezione di Torino, via P. Giuria, 1, I-10125 Torino, Italy}
\date{January,23 2007}
\newcommand{\eq}{\begin{equation}}
\newcommand{\en}{\end{equation}}
\newcommand{\ear}{\begin{eqnarray}}
\newcommand{\rae}{\end{eqnarray}}
\newcommand{\Z}{\mathbb{Z}}
\newcommand{\C}{\mathbb{C}}
\newcommand{\uu}{\mathbb{I}}
\newcommand{\R}{{\cal R}}
\newcommand{\s}{{\cal S}}
\newcommand{\T}{{\cal T}}
\newcommand{\D}{{\cal D}}
\newcommand{\N}{{\cal N}}
\newcommand{\bra}{\langle}
\newcommand{\ket}{\rangle}
\newcommand{\um}{\frac12}
\newcommand{\tr}{{\rm tr}\,}
\newcommand{\Tr}{{\rm tr}^{~}}

\begin{abstract}
The blackbody radiation  in a box 
$L^3$ with periodic boundary conditions in thermal equilibrium at a 
temperature $T$ is affected by  finite-size effects. These bring about
modifications of the thermodynamic functions which can be expressed 
in a closed form in terms of the dimensionless
parameter $L\,T$.  For instance, when $L\,T\sim 4$ - corresponding to the 
value where the most reliable $SU(N)$ gauge lattice simulations have been 
performed above the deconfining temperature $T_c$ - the deviation of the 
free energy density from its thermodynamic 
limit is about 5\%. This may account for almost half of the pressure 
deficit observed in lattice simulations at $T\sim 4\, T_c$.   
\end{abstract}
\pacs{11.10.Wx; 11.15.Ha; 12.38.Mh}
\maketitle
In a very hot quark gluon plasma, when the temperature  is much larger than 
any other relevant mass scale, asymptotic freedom leads  to expect that the effective coupling to be used in thermodynamical calculations should be small. However, even in the case where the coupling is very 
small, strict perturbation theory cannot be used, the reason being that 
infrared divergences occur in high order calculations and various 
resummations are needed to get meaningful results. Lot of effort has been 
devoted  to the perturbative calculations of the 
pressure \cite{az,bn,ml1,ml2,ml3}.
The values obtained by adding successively high order contributions 
oscillate too much, and strongly depend on the renormalization scale. 
Thus such a plasma cannot be described simply as a gas of weakly 
interacting quarks and gluons.

 This very conclusion was also reached by lattice studies both
 for pure $SU(3)$ gauge case \cite{bb} and for different 
kinds of fermions \cite{en,Karsch:2000ps,Bernard:2006nj}. 
Such calculations revealed a slow approach to 
the ideal gas limit of the thermodynamic functions. 
In particular it resulted a large deficit in the pressure 
and entropy as compared to the Stefan-Boltzmann 
law for free gluon gas, which 
remained at the level of more than 10\% even at temperatures as high as
$T\simeq4\,T_c$. Similar results have also been found for $SU(4)$ and 
$SU(8)$ gauge theories in a more restricted range of temperatures \cite{bt}.

These simulations were made on lattices of size $N_s^3\times
 N_t$ with periodic boundary conditions. Much effort has been dedicated to  
study and  control  the ultraviolet (UV) cut-off effects which are in general 
$O((aT)^2) $. In the standard Wilson formulation, temporal extent 
$N_t=1/aT \ge 8$ is needed in order to get  reliable extrapolations of 
the thermodynamic functions  to the continuum limit. 

In this paper I wish to focus on another facet of lattice simulations, 
i.e. the infrared (IR) finite-size effects. In fact, for a thorough 
comparison of the numerical data of the hot quark-gluon plasma with the 
Stefan-Boltzmann (SB) law, one should consider a free gluon gas enclosed 
in a box with the same size and the same boundary conditions 
of the corresponding numerical experiment.  
It is clear that finite size effects are expected to be particularly 
relevant in a free boson gas: the lack of an intrinsic length 
scale leads to a maximal sensitivity to the geometrical shape $N_s/N_t=LT$ 
of the system. The purpose of this paper is to evaluate these 
infrared effects. 

This might appear to be an academic exercise in view of the 
fact that in  $T\to\infty$  limit the internal 
energy density $\epsilon$ of the resulting ideal 
lattice gluon gas has been explicitly evaluated \cite{bkl} 
through numerical 
integration both in the thermodynamic limit and for fixed $N_s/N_t$.
The finite-size effects turn out to be of the order of 1\% for $N_s/N_t=4$   
\cite{bkl} (see Fig.1). Moreover, in the continuum limit this system is scale
invariant, therefore  the trace of the energy-momentum tensor is vanishing
\eq
T^\mu_\mu\equiv\epsilon-3\,p=0~~ (T\to\infty,a\to0,N_s/N_t=const)~, 
\label{tmumu}
\en
hence the deviations of the pressure from the SB value are even 
smaller.

There is however a missing  point in the above reasoning. The 
quantity which is calculated in lattice simulations is not  the 
pressure $p$, but the free energy density $f$, the reason \cite{efk} 
being that the evaluation of $p$ would involve the derivative of the 
bare coupling with respect to the volume which is known only perturbatively, 
while $f$ can be evaluated in a sounder way and in the thermodynamic limit 
one has $p=-f$. In a finite volume, however, this relation is violated. 
A numerical study on a $SU(2)$ gauge system at intermediate temperatures  
suggested the 
empirical rule $L_s/L_t\gtrsim4$ \cite{efk} to get rid of the finite-size 
corrections. In fact this bound is not enough at high temperature: in the 
free gluon gas limit, where the canonical partition function $Z$ 
can be evaluated exactly even on finite volume, we shall prove that
\eq
\frac{\log Z}{N^2-1}=\frac{\pi^2}{45}(LT)^3-\log\sqrt{LT}+
O(e^{-2\pi\,LT})~.
\label{logz}
\en 
As expected, this quantity is not purely extensive, owing to the finiteness 
of the  volume $V=L^3$. Its deviation from the thermodynamic limit is 
a universal function of  $LT$. We may derive  from $Z$  
 any other thermodynamical 
function. The  quantities we will concentrate on are  
the pressure $p$, the internal energy density $\epsilon$ and the free energy 
density. These are given by, neglecting  exponential corrections, 
\eq
p\equiv T\left(\frac{\partial\log Z}{\partial V}\right)_T=
(N^2-1)\left(\frac{\pi^2T^4}{45}-\frac T{6\,V}\right)\,,
\en
\eq
\epsilon\equiv\frac{T^2}{V}\left(\frac{\partial\log Z}{\partial T}\right)_V=
(N^2-1)\left(\frac{\pi^2T^4}{15}-\frac T{2\,V}\right)\,,
\label{energy}
\en

\eq
-f\equiv\frac{T}{V}\log Z=
(N^2-1)\left(\frac{\pi^2T^4}{45}-\frac{T}{V}\log\sqrt{LT}\right)\,.
\en
According to (\ref{tmumu}) we have $\epsilon-3p=0$, while
\eq
\frac{p+f}{T^4}=\frac{N^2-1}6\frac{\log(LT)^3-1}{(LT)^3}\,,
\label{pf}
\en
therefore we cannot trade $p$ for $-f$ unless $LT$ is  large enough.
The presence of a log term in $f$  makes it vary  very slowly with the
shape $LT$ (see Fig.2). The variation is almost 10\% for $N_s/N_t\simeq 3$
where the simulations with staggered quarks are currently performed 
\cite{Bernard:2006nj} and reduces to 1\% only at $N_s/N_t\simeq 8$ which 
seems presently unattainable.

\begin{figure}[tb]
\begin{center}
\psfrag{e/e}{$\frac{\epsilon(N_t/N_s,N_t)}{\epsilon(0,N_t)}$}
\psfrag{a}{{\color{blue}$\frac{\epsilon(0,N_t)}{\epsilon_{SB}}$}}
\psfrag{n}{$N_t$}
\psfrag{r}{$\frac{N_s}{N_t}=6$}
\psfrag{s}{$\frac{N_s}{N_t}=4$}
\mbox{~\epsfig{file=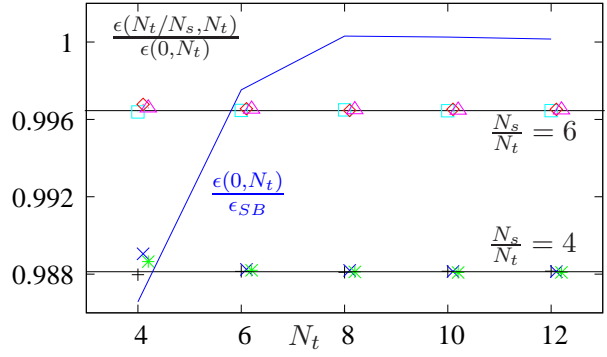,width=8.cm}}
\vskip .2 cm
\caption{Deviations from the thermodynamic limit ($N_s\to\infty$) of 
three different ideal lattice gases with finite spatial extent for two
different values of the ratio $N_s/N_t$. Some symbols are shifted to the 
right for clarity. For comparison, the energy density of one of 
these models is drawn in units 
of the continuum limit to show the dependence on the cut-off. The data are 
taken from Table I and II of Ref.\cite{bkl}. One 
observes a remarkable cancellation of the UV cut-off effects for $N_t>4$ 
and the emergence of a universal function of  $\frac{N_s}{N_t}$. 
 The horizontal lines are 
not fits, but denote the two values assumed by the function 
$1-\frac{15}{2\pi^2}(N_t/N_s)^3$ as predicted by Eq. (\ref{energy}).}
\label{Figure:1}  
\end{center}
\end{figure}

\begin{figure}[tb]
\begin{center}
\psfrag{ff}{$f/f_{\rm SB}$}
\psfrag{nsnt}{$N_s/N_t$}
\psfrag{formula}{{
\color{blue}$1-\frac{45\,N_t^3}{2\pi^2N_s^3}\log\frac{N_s}{N_t}$}}
\mbox{~\epsfig{file=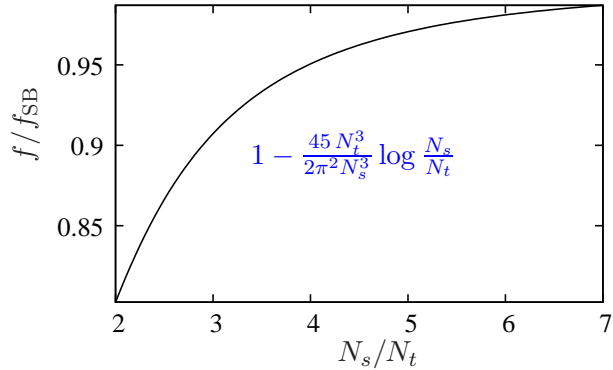,width=8.cm}}
\vskip .2 cm
\caption{Deviation from the thermodynamic limit $f_{\rm SB}$  of 
the free energy density $f$ of the free gluon gas.}
\label{Figure:2}  
\end{center}
\end{figure}

A non-trivial numerical check of the above formulae comes from 
the mentioned study \cite{bkl} on 
$T\to \infty$ limit of $SU(3)$ gauge theory, where  three different $O(a^2)$ 
and $O(a^4)$ improved actions were used to explore finite-size effects on
 ideal lattice gluon gases. The energy density $\epsilon(N_t/N_s,N_t)$ 
was evaluated by numerical integration for some values of $N_t$ 
and for two different ratios $N_s/N_t=4$ and $N_s/N_t=6$ (5 significant digits)
as well as in the thermodynamic limit $\epsilon(0,N_t)$ (7 significant digits).
Their ratios are plotted in Fig.1 as functions of $N_t$. In principle, these 
 should be model-dependent functions of $N_s/N_t$ {\sl and} $N_t$, 
but there is a remarkable cancellation of the UV cut-off dependence 
for $N_t>4$. As a consequence, there is no other intrinsic length scale in 
the game besides the size of the system, hence these ratios are expected 
to collapse toward a universal function of $N_s/N_t$ which, according to 
(\ref{energy}), should be $1-\frac{15}{2\pi^2}(N_t/N_s)^3$. The numerical data
fit perfectly this prediction within  numerical accuracy.
  
Let us come to a proof of the main formula (\ref{logz}). 
For sake of generality, I shall treat the case of a massless scalar field $\phi$ in a (hyper)cubic box of volume $V=L^D$ at a temperature $T$. The canonical 
partition function $Z$ is defined by a functional integral over all  
periodic field configurations
with period $L$ in the space directions and with period 
$\beta=1/T$ in the imaginary time. The dimensions of the box 
are large compared to $\beta$, i.e. $LT\gg1$.

There is a rich 
literature on a strictly related subject, the Casimir effect, where different 
methods have been developed to study various kinds of finite-size effects
(see e.g. \cite{Edery:2005bx} for useful formulae and further references).
It is however more instructive to account for the functional form of $Z$ 
by exploiting some simple symmetry principles. First, we require that $Z$ be dimensionless, of course. Owing to the absence of intrinsic length scales, $Z$ should be a function of the unique dimensionless parameter $LT$. This yields 
invariance under the scale transformation 
\eq
Z=Z(LT)\Leftrightarrow ~~L\to s\,L~,\beta\to s\,\beta
\en
Secondly, the $D+1$ periodic box defines a cell of an infinite, regular, 
lattice.
The physics should not depend on the choice of the fundamental region 
tiling  the whole $D+1$-dimensional space by discrete translations. 
This requires  modular invariance of the system. 

Periodic boundary conditions allow for a zero mode 
$\phi=\phi_o$ of the scalar field. In a system with zero modes  
the functional integral  splits into two factors
\eq
Z(LT)=C_o \,\D=\,C_o\,[{\rm Det} K]^{-\um}~,
\label{split}
\en
where $C_o$ denotes the  zero mode contribution  while the other factor
is the integral 
over the Gaussian fluctuations around the zero modes, described by 
the kinetic operator $K$. The latter can be 
written  as the product over the  eigenvalues $\lambda_k$ of $K$ 
 in the usual form 
\eq 
[{\rm Det} K]^{-\um}=\prod_k'\lambda_k^{-\um}~,
\en
the prime here indicating that we are to exclude the zero eigenvalue.
Under general grounds (see e.g.\cite{do} or \cite{we}, page 463) one can prove that a 
rescaling $\lambda_k\to s^{-2}\lambda_k$
 of all  non-vanishing eigenvalues yields correspondingly 
 $    [{\rm Det} K]^{-\um}\to s^{\N}[{\rm Det} K]^{-\um} $, 
where $\N$ denotes the number of zero modes.
In the present case $K=-\partial^2$ where $\partial^2$ is the Laplacian, and
\eq
\lambda(m_i,n)=\sum_{i=1}^D\left(\frac{2\pi m_i}{L}\right)^2+\left(\frac{2\pi n}{\beta}\right)^2
\en  
where the $m_i$'s and $n$ run from $-\infty$ to $\infty$. As a consequence, 
$\D\equiv[{\rm Det} -\partial^2]^{-\um}$ is a function of $L$ and $\beta$  
 and the mentioned rescaling property of $\D$ becomes
\eq
\D(s\,L,s\,\beta)= \,s\,\D(L,\beta)~,
\en
showing that the zero mode  factorization (\ref{split}) spoils the 
scale invariance of $\D$. In order to recover the scale 
invariance of $Z$, the factor $C_o$ 
should have dimensions of  length. The only modular invariant quantity with 
this property can be constructed with the volume of the $D+1$-dimensional 
cell, namely, $(L^D\beta)^{1/(D+1)}$. Choosing $s=1/L$ we get, aside from 
irrelevant numerical factors,
\eq
Z=(\beta/L)^{1/(D+1)}\,\D(1,\beta/L)~.
\label{zp}
\en
To  explicitly evaluate $\D$ it is convenient to resort to the $\zeta$-function
regularization \cite{rs,ha,dc} of the Laplacian determinant
\eq
  \D(1,\beta/L) =\prod_{m_i,n}' \left(\sum_{i=1}^D m_i^2+  (LTn)^2\right)^{-\um}~.
\label{det}
\en
where the prime now indicates a regularised product.
One of the virtues of the $\zeta$ function regularization is that one can 
deal with regularised sums or products as they were absolutely convergent 
series and products
\cite{df,Gliozzi:1992wa,Gliozzi:2005cd}. In (\ref{det}) we consider only 
the factors with $n\not=0$ 
because the others generate irrelevant numerical constants. It is useful 
to classify the factors according to the number $k$ of non vanishing $m_i$. 
Denoting with $c_k$ the product over the set 
$\{m_1\not=0,\dots m_k\not=0,n\not=0\}$ and with $p_k$ the product  over
the set $I_k=\{m_1>0,\dots m_k>0,n>0\}$
we have $ c_k=p_k^{2^{k+1}}$ with
\eq
p_k=\prod_{I_k}'
\left[\sum_{i=1}^k m_i^2+  (LTn)^2\right]~,
\en
and

\eq
\log\D(1,\beta/L)=-\sum_{k=0}^D
\left(
\matrix{
D\cr
k}
\right)2^{k}\log\,p_k~.
\label{dp}
\en
We now use the $\zeta$-regulated product $\prod_{m>0}'m^2=2\pi$ to
rewrite $p_k$ in the form
\eq
p_k=\prod_{I_{k-1}}'2\pi\prod_{m>0}
\left[1+\frac{\sum_{i=1}^{k-1}m_i^2+(LTn)^2}{m^2}\right]~.
\en 
According to the known formula
\eq
\prod_{m>0}(1+\frac{\alpha^2}{m^2})=
\frac{e^{\pi\alpha}}{2\pi\alpha}(1-e^{-2\pi\alpha})~,
\en
we are led to the key identity
\eq
p_{k+1}=p_{k}^{-\um}\,e^{L\,E_C^{(k)}}\,Q_{k}
\label{key}
\en
with
\eq
Q_k=\prod_{I_{k}}\left[1-e^{-2\pi
\sqrt{\sum_{i=1}^{k}m_i^2+(LTn)^2}}\right]~.
\en
The $\zeta-$regulated quantity
\eq
E_C^{(k)}=\um\sum_{I_k}'\sqrt{\sum_{i=1}^{k}\left(\frac{2\pi m_i}{L}\right)^2+
\left(\frac{2\pi n}\beta\right)^2}
\en
is the Casimir energy of $\phi$ in  a box of size 
$\frac\beta2\times(\frac{L}2)^k$
with Dirichlet boundary conditions. Inserting (\ref{key}) in (\ref{dp}), 
these Casimir energies combine to form the quantity
\eq
E_o=\sum_{k=0}^{D-1}2^k\left(\matrix{D-1\cr k}\right)\,E_C^{(k)}~.
\en
This is the zero-point energy of the massless scalar field in the 
{\sl periodic} box of size $\beta\times L^{D-1}$, which is exactly known 
even at finite values of $LT$ (see e.g.\cite{Edery:2005bx}):
\eq
E_o=-\frac{L^{D-1}}{\beta^D}\frac{\Gamma(\frac{D+1}2)}{\pi^{\frac{D+1}2}}
\,\zeta(D+1)~,
\en
where $\zeta(x)$ is the Riemann zeta function. 
Applying this to Eq. (\ref{dp}) and defining 
\eq
\log Q=\sum_{k=0}^{D-1}2^{k+1}\left(\matrix{D-1\cr k}\right)\,\log\, Q_k
\simeq e^{-2\pi LT}+\dots
\en
we may rewrite Eq.(\ref{zp}) in the form
\eq
\log Z=\frac{L^{D}}{\beta^D}
\frac{\Gamma(\frac{D+1}2)}{\pi^{\frac{D+1}2}}\zeta(D+1)-
\frac{\log\frac L\beta}{D+1}
-\log Q\,.
\en
This is the final result.
It can be rewritten in a more evocative form  
\eq
\log Z=-L\,E_o- \frac{\log LT}{D+1} 
-\sum_{\vec{k}}\log(1-e^{-L\omega_{\vec{k}}})~,
\en
where $Z$ can now be viewed as the canonical partition function of 
$\phi$ in the asymmetric, periodic, box $L^{D-1}\beta$ in equilibrium at the 
``temperature'' $1/L$; the sum is over the momenta of the normal modes of 
energy $\omega_{\vec k}$. The rotation, or modular transformation, of the 
fundamental $D+1$ cell with respect to the standard approach to blackbody 
radiation has made it possible to highlight the finite size effects of
$\log Z$.
 Aside from small exponential corrections, it 
differs from the  thermodynamic limit by a non-negligible 
logarithmic term. The latter is essential to enforce modular invariance, 
which was obvious at the beginning of the calculation. Its origin 
can be traced to the zero mode subtraction in Eq.(\ref{split}).

When $D=1$ we recover the known partition  function of a scalar massless 
boson on a 2-torus. Of course, for a free gluon gas 
one has to multiply this result by the number of polarisation states 
and by the dimensions of the adjoint representation of the gauge group.

In conclusion, in this paper it has been pointed out that the free energy 
of an ideal gluon gas in a periodic box is affected by  non-negligible IR 
finite-size effects. It would be  very interesting to observe similar effects 
in the interacting case. Unfortunately, in order to see them 
varying the ratio $T/T_c$ by acting on the coupling constant does not suffice:
one has to modify the ratio $N_s/N_t=LT$. Notice that the finite-size 
deviation of $f$ does not influence only the evaluation of the pressure 
but also the internal energy  and the entropy. In fact in lattice simulations 
one calculates two different physical quantities: the trace anomaly 
$\Delta=\epsilon-3p$ and the free energy density $f$, and one defines the
energy density as $\epsilon\simeq \Delta-3f$ and the entropy density 
as $s\simeq(\Delta-4f)/T$; in view of Eq.(\ref{pf})
 these are true only in the thermodynamic limit. It has been observed 
\cite{bb} that the cut-off dependence in $\Delta$ is much smaller than 
for the pressure alone. This could be a clue to IR finite-size effects.      

\end{document}